\UseRawInputEncoding
\documentclass[aps,floatfix,amsmath,nofootinbib,amssymb,superscriptaddress]{revtex4}

\usepackage{overpic}
\usepackage{amssymb}
\usepackage{indentfirst}
\usepackage{feynmf}   
\usepackage{slashed}  
\usepackage{cases}
\usepackage{color}
\usepackage{multirow}
\usepackage{epstopdf}
\usepackage{graphicx,color,bm}
\usepackage{epstopdf}

\usepackage[colorlinks,
            citecolor=blue,
            anchorcolor=red,
            menucolor=red,
            linkcolor=red,
            filecolor=red,
            runcolor=red,
            urlcolor=blue,
            frenchlinks=red]{hyperref}

\begin{document}

\title{Weighted $\sin\left(3\phi_h -\phi_S\right)$ asymmetry from pretzelosity in SIDIS at electron ion colliders}

\author{Shi-Chen Xue}
\affiliation{School of Physics and Microelectronics, Zhengzhou University, Zhengzhou, Henan 450001, China}
\author{Xiaoyu Wang}
\email{xiaoyuwang@zzu.edu.cn}
\affiliation{School of Physics and Microelectronics, Zhengzhou University, Zhengzhou, Henan 450001, China}
\author{De-Min Li}
\email{lidm@zzu.edu.cn}
\affiliation{School of Physics and Microelectronics, Zhengzhou University, Zhengzhou, Henan 450001, China}
\author{Zhun Lu}
\email{zhunlu@seu.edu.cn}
\affiliation{School of Physics, Southeast University, Nanjing 211189, China}

\begin{abstract}
We investigate the transverse momentum weighted azimuthal asymmetry with a $\sin\left(3\phi_h -\phi_S\right)$ modulation in electroproduction of charged pion production off the transversely polarized proton.
The asymmetry is contributed by the product of the second transverse-moment of the pretzelosity distribution function and the first transverse-moment of the Collins function.
Using the extracted result for the proton pretzelosity and the available parametrization for the pion Collins function, we numerically predict the $P_{hT}^3$-weighted $\sin\left(3\phi_h -\phi_S\right)$ asymmetry for charged pion production at electron ion colliders (EICs).
It is found that the weighted asymmetry is around  2 percent in the kinematics region of EICs.
Our results show that the planned high luminosity electron ion colliders could be feasible to access pretzelosity in the near future.
\end{abstract}

\maketitle

\section{INTRODUCTION}

The question how the nucleon spin is distributed among its constituents is always a main challenge in QCD spin physics~\cite{Jaffe:1989jz,Ji:1996ek,Deka:2013zha,Sato:2016tuz,Ethier:2017zbq,Yang:2016plb,Kunne:2019hjt}.
The chiral-odd pretzelosity distribution $h_{1T}^\perp$~\cite{Bandos:1995dw,Miller:2007ae,Miller:2003sa},
the least known distribution among the eight leading-twist transverse momentum dependent (TMD) distribution functions, describes the density of quarks transversely polarized along the $y$ axis, in
a nucleon that is transversely polarised along the $x$ axis.
Thus, it is closely related to the non-spherical nucleon shape.
Particularly, it is found that the pretzelosity distribution and the orbital angular momentum (OAM) of quarks inside the nucleon satisfy the following model inspired relation~\cite{She:2009jq,Avakian:2010br,Efremov:2010cy,Lorce:2011kn}:
\begin{equation}
{\cal L}_{z}^a = -\int d x \,  d^2 \bm{p}_{T} \, \frac{p_T^2}{2M^2} h_{1T}^{\perp a}(x, p_T
^2) = -\int d x  \; h_{1T}^{\perp (1) a}(x) \;.
\label{oam}
\end{equation}
Therefore, studying pretzelosity will provide an important cross-check on quark OAM and will be an important ingredient toward the full understanding of the nucleon spin structure.

In recent years, the pretzelosity distribution has been intensively studied by different models~\cite{Avakian:2008dz,Pasquini:2008ax,She:2009jq,Avakian:2010br,Lorce:2011kn,Zhang:2013dow,
Lefky:2014eia,Liu:2014zla,Gutierrez-Reyes:2017glx,Gutierrez-Reyes:2018iod,Chai:2018mwx}.
Several QCD-inspired quark models have been adopted to calculate pretzelosity, such as the light-cone quark model~\cite{Pasquini:2008ax}, the diquark model~\cite{Bacchetta:2008af,She:2009jq}, and the bag model~\cite{Avakian:2008dz,Avakian:2010br}.
These calculations provide a first glimpse on the size and sign of pretzelosity of valence quarks.
On the other hand, various spin-dependent azimuthal asymmetries in high energy scattering processes, such as the semi-inclusive deeply inelastic scattering~(SIDIS)~\cite{Bacchetta:2006tn} and dilepton production in hadron-nucleon collisions~\cite{Arnold:2008kf} (Drell-Yan), can be applied to explore the internal structure of the nucleon.
Recently, the authors in Ref.~\cite{Lefky:2014eia} extracted the pretzelosity of the up and down quarks, from SIDIS data on the $\sin\left(3\phi_h -\phi_S\right)$ azimuthal asymmetry.
For the first time, the extrated results show tendency for up-quark pretzelosity to be positive and down-quark pretzelosity to be negative.
Due to limited statistics of experimental data, there are still large uncertainties on the values of parameters.
In Ref.~\cite{Bastami:2020asv}, based on the model results and the parametrization for pretzelosity, the $\sin(2\phi+\phi_S)$ azimuthal asymmetry, which is due to the convolution of pretzelosity and pion Boer-Mulders function, was estimated and compared with the COMPASS data.

As the contribution from pretzelosity is proportional to $P_{T}^3$, with $P_{T}$ being the transverse momentum of the particle observed in final state, it will lead to a rather small asymmetry due to the kinematic suppression.
Furthermore, to survive in a high energy scattering process, another chiral-odd distribution/fragmentation function is needed to couple with pretzelosity due to its chiral-odd nature.
These features make it quite difficult to measure pretzelosity experimentally.
Thus, it is necessary to study the kinematical region which is sensible to the effect of pretzelosity.
For this purpose, in this work we will study the feasibility to probe pretzelosity in the high luminosity SIDIS experiments such as the electron ion colliders.
The direct observable in SIDIS involving pretzelosity in leading-twist is the structure function $F_{UT}^{\sin\left(3\phi_h -\phi_S\right)}$.
It appears in the electroproduction of unpolarized hadron off transversely polarized nucleon target,
and is expressed as the convolution of the pretzelosity and the Collins function $H_1^\perp$~\cite{Collins:1992kk}, the latter one describes the fragmentation of a transversely polarized quark to an unpolarized hadron and
can be accessed from the azimuthal asymmetries of two back-to-back hadron productions in $e^+ e^-$ annihilations~\cite{Boer:1997mf,Seidl:2008xc,TheBABAR:2013yha}.

There are two types of commonly studied azimuthal asymmetries: transverse momentum weighted and unweighted.
In this work, we choose to estimate the weighted $\sin\left(3\phi_h -\phi_S\right)$ asymmetry.
From the theoretical aspects, weighted asymmetry has the advantage that it can be expressed as the product of collinear objects instead of complicated convolution in the transverse momentum space. In the case of weighted $\sin\left(3\phi_h -\phi_S\right)$ asymmetry the second transverse moment of pretzelosity $h_{1T}^{\perp (2)}(x)$ and the first transverse moment of the Collins function $H_1^{\perp (1)}(z)$ are involved.
As there is no transverse momentum in these objects, they allow a simpler study of scale evolution and could be extracted from the data of weighted asymmetry in a model-independent way.
However, from the experimental side, the weighted asymmetry is more difficult to measure since high statistics is needed.
With the advantages of high luminosity and collider mode, the future EIC in the US~\cite{Accardi:2012qut} and electron ion collider in China (EicC)~\cite{Cao:2020} could provide more stringent constraints on the $\sin\left(3\phi_h -\phi_S\right)$ asymmetry with reasonable accuracy.

The rest of the paper is organized as follows.
In Sec.~\ref{sec:formalisms}, we provide the theoretical expression of the $P_{hT}^3$-weighted $\sin\left(3\phi_h -\phi_S\right)$ asymmetry of charged pion production in SIDIS process.
In Sec.~\ref{sec:numerical}, we estimate the weighted asymmetry at the kinematics of the EIC and EicC using the available parametrization for the pion Collins function and the proton pretzelosity function as inputs.
In the calculation we consider the DGLAP evolution of the collinear objects. We summarize the paper and discuss the results in Sec.~\ref{sec:conclusion}.

\section{FORMALISM OF THE $\mathbf{P_{hT}^3}$-weighted $\sin\left(3\phi_h -\phi_S\right)$ ASYMMETRY IN SIDIS PROCESS}
\label{sec:formalisms}

\begin{figure}
  \centering
  \includegraphics[width=0.49\columnwidth]{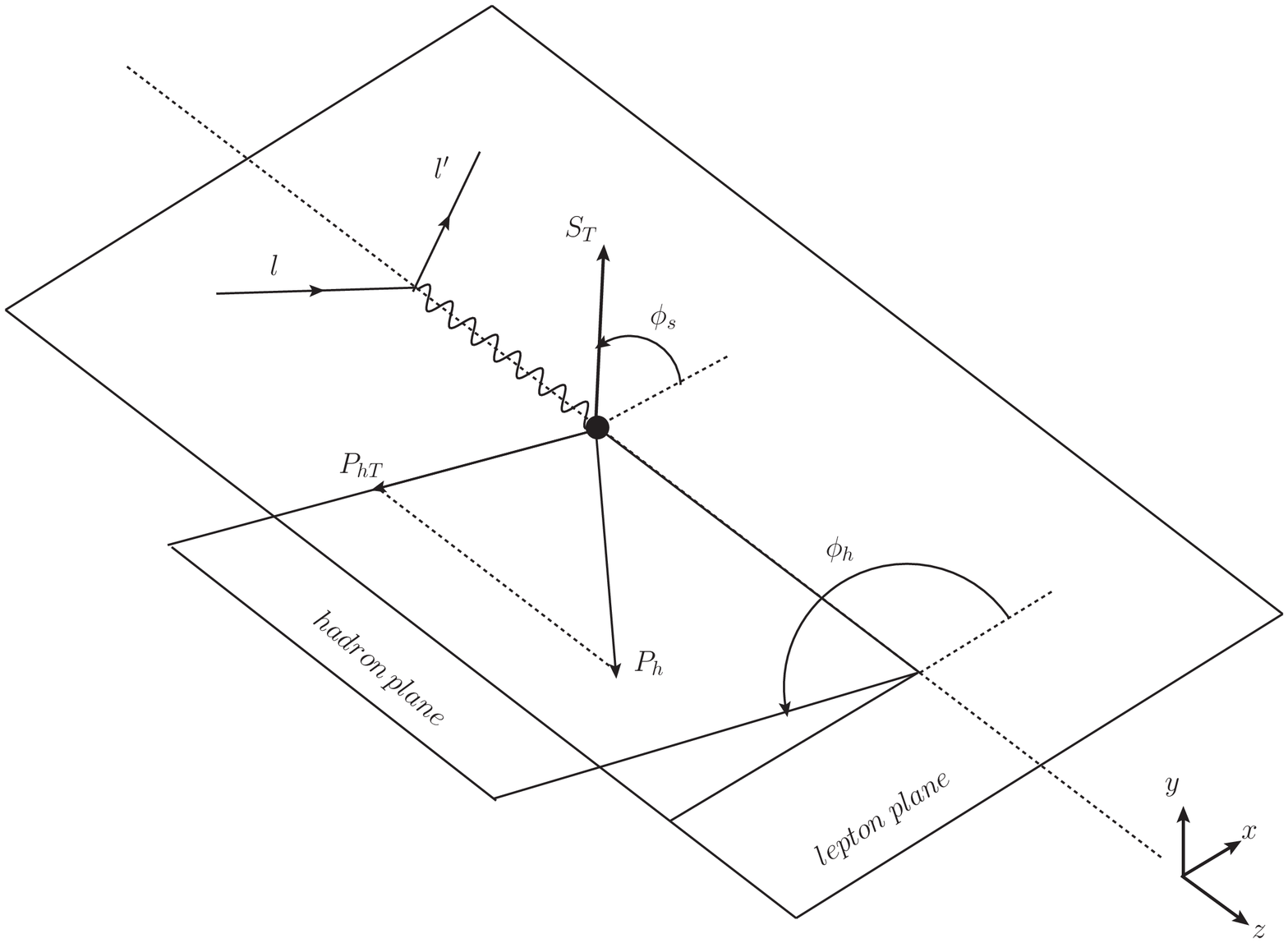}
  \caption{Reference frame of the transversely polarized SIDIS~\cite{Bacchetta:2004jz}.  }
  \label{lhp}
\end{figure}

The process we are interested in is the transversely polarized SIDIS:
\begin{equation}
\label{eq:sidis}
e(\ell)+p^\uparrow(P) \longrightarrow e(\ell^\prime)+\pi (P_\pi)+X(P_X),
\end{equation}
where $e$ denotes the electron beam, $p$ is the proton target, $\pi$ is the semi-inclusively produced hadron, and $\uparrow$ represents that the proton target is transverse polarized.
The reference frame of the process under study is shown in Fig.~\ref{lhp}, in which the momentum direction of the virtual photon defines the $z-$axis according to the Trento conventions~\cite{Bacchetta:2004jz}.
$P_{hT}$ and $S_T$ are the transverse component of $P_h$ and the spin vector $S$, respectively.
$\phi_h$ denotes the the azimuthal angle of the final hadron around the virtual photon, and $\phi_S$ stands for the angle between the lepton scattering plane and the direction of the transverse spin of the nucleon target.

We introduce the following invariants to express the differential cross section of the process as well as the experimental observables
\begin{eqnarray}
x&=&\frac {Q^2}{2P \cdot q}\,, \quad  y=\frac{P \cdot q}{P \cdot \ell}=\frac {Q^2}{x s}\,,\quad z=\frac{P \cdot P_h}{P \cdot q}\,,\nonumber \\
Q^2&=&-q^2\,,\quad\quad s=(P+\ell)^2\,,\quad \quad\;\gamma = \frac {2Mx}{Q} ,
\end{eqnarray}
where $s$ is the center of mass energy squared of the $e-p$ system,
$q=\ell-\ell^\prime$ denotes the momentum of the virtual photon with invariant mass squared $Q^2=-q^2$.

Assuming single-photon exchange, the SIDIS cross section can be expressed in terms of 18 structure functions~\cite{Bacchetta:2006tn} in a model-independent way.
Here, we only consider the $\sin\left(3\phi_h -\phi_S\right)$ modulation and neglect other unrelated terms
\begin{eqnarray}
\frac{d\sigma}{dx dy dz d\phi_S d\phi_h d\bm{P}^2_{hT}} &=&
\frac{\alpha^2}{s}\,\left( 1+\frac{\gamma^2}{2x}\right)\,
\biggl[A(x,y)\,F_{UU} \nonumber\\
&+& |\bm{S}_{\perp}|\,B(x,y)\,\sin (3\phi_h-\phi_S)\,F_{UT}^{\sin (3\phi_h-\phi_S)} + \ldots\biggr], \label{eq:cs}
\end{eqnarray}
where $\alpha=e^2/(4\pi)$ is the fine structure constant, and the depolarization factors have the form
\begin{eqnarray}
A(x,y) &=& \frac{1}{x^2 y^2\, (1+\gamma^2)} \,
\left( 1-y+ \frac{1}{2}y^2 + \frac{1}{4} y^2 \gamma^2 \right) \; ,  \\
B(x,y) &=& \frac{1}{x^2 y^2 \, (1+\gamma^2)} \,
\left( 1-y - \frac{1}{4} y^2 \gamma^2 \right) \;  .
\end{eqnarray}
In Eq.~(\ref{eq:cs}), $F_{UU}$ stands for the unpolarized structure function and $F_{UT}^{\sin\left(3\phi_h -\phi_S\right)}$ is the transverse spin-dependent structure function, with the subscripts $U$ or $T$ denoting the polarization states of the beam (first subscript) and the target (second subscript).
Using the notation $\mathcal{C}$,
\begin{equation}
\label{eq:note_C}
\mathcal {C}\bigl[ \omega f  D \bigr]
= x\sum_q e_q^2 \int d^2 \bm{p}_T  d^2 \bm{k}_T \delta^{(2)}\bigl(\bm{p}_T - \bm{k}_T - \bm{P}_{hT}/z \bigr)\omega(\bm{p}_T,\bm{k}_T)
f^q(x,p_T^2)\,D^q(z,k_T^2),
\end{equation}
the two structure functions can be expressed as the following convolutions~\cite{Bacchetta:2006tn}:
\begin{eqnarray}
F_{UU} &=& {\cal C} \biggl[ f_1 D_1 \biggr] \; ,  \\
F_{UT}^{\sin (3\phi_h-\phi_S)} &=& {\cal C} \biggl[
\frac{2 (\hat{\bm{h}}\cdot \bm{p}_{T})\,(\bm{k}_{T}\cdot \bm{p}_{T})+
      \bm{p}_{T}^2\,(\hat{\bm{h}}\cdot \bm{k}_{T})
      -4 (\hat{\bm{h}}\cdot \bm{p}_{T})^2\,(\hat{\bm{h}}\cdot \bm{k}_{T})}
     {2 M^2 M_h} \, h_{1T}^{\perp}\,H_1^{\perp} \biggr] \; , \label{eq:conv_ut}
\end{eqnarray}
where $\hat{\bm{h}} = \bm{P}_{hT} / |\bm{P}_{hT}|$, $f_1(x,\bm{p}_T)$ and $D_1(z,\bm{k}_T))$ are the unpolarized TMD distribution and fragmentation function, respectively.
From Eq.~(\ref{eq:conv_ut}), one can see that the coupling of the pretzelosity distribution function of the nucleon target and the Collins function of the final-state pion meson contributes to the $\sin\left(3\phi_h -\phi_S\right)$ azimuthal asymmetry.

In order to simplify the complicated TMD evolution effects as well as the convolution in the transverse momentum space, we will consider the transverse momentum weighted asymmetries,
which are usually expressed as the simple product of the moments of the TMD distribution/fragmentation functions instead of introducing  model dependence on transverse momenta.
In SIDIS process, the $P_{hT}^3$-weighted asymmetries are defined as~\cite{Boer:1997nt,Bacchetta:2010si}
\begin{equation}
A_{XY}^W(x,y,z) \propto \frac{\langle W \rangle_{XY}}{\langle 1 \rangle_{UU}}
\equiv \frac{\int d\phi_S d\phi_h d\bm{P}^2_{hT}\,W\, d\sigma_{XY}}
            {\int d\phi_S d\phi_h d\bm{P}^2_{hT}\,d\sigma_{UU}} \; ,
\label{eq:wasymmdef}
\end{equation}
where $d\sigma_{XY}$ denotes the differential cross section for lepton beam with polarization $X$ and proton target with polarization $Y$,
$W$ represents a function of the azimuthal angles $\phi_h$, $\phi_S$ and suitable powers of $Q_T=|\bm{P}_{hT}|/z$.
In this study, we choose the weighting function to be $Q_T^3\,\sin (3\phi_h-\phi_S)$, which leads to the following weighted $\sin\left(3\phi_h -\phi_S\right)$ asymmetry~\cite{Bacchetta:2010si}
\begin{eqnarray}
A_{UT}^{Q_T^3\,\sin (3\phi_h-\phi_S)} &= 2\,
\frac{\Bigl\langle \frac{Q_T^3}{6M^2 M_h}\, \sin (3\phi_h-\phi_S) \Bigr\rangle_{UT}}
     {\langle 1 \rangle_{UU}} \stackrel{\rm TMD}{=}  2 \,
\frac{\sum\limits_a e_a^2 \, x h_{1T}^{\perp (2) a}(x)\,H_1^{\perp (1) a}(z)}
     {\sum\limits_a e_a^2\, x f_1^a(x)\,D_1^a(z)} \; ,
\end{eqnarray}
where ${h}^{\perp(2)a}_{1T}(x)$ is the second transverse-moment of pretzelosity,
and $H_1^{\perp (1) a}(z)$ is the first transverse-moment of the Collins function:
\begin{eqnarray}
{h}^{\perp (2)a}_{1T}(x)=\int {d}^{2}\bm{p}_T\left(\frac{\bm{p}_T^2}{2M_\pi^2} \right)^2 {h}^{\perp a}_{1T}(x,\bm{p}_T^2),
\label{eq:moment}\\
H_1^{\perp (1) a}(z) = \int d^2 {\bm{k}}_{T} \ \frac{|k_T|^2}{2z^2M_h^2}  H_1^{\perp a}(z,\bm{k}_T ^2)\; .\
\label{eq:moment2}
\end{eqnarray}
Furthermore, in the Trento convention~\cite{Bacchetta:2004jz}, the first transverse moment $H_1^{\perp (1) }(z)$ can be related to the twist-3 collinear fragmentation function $\hat H(z)$ via~\cite{Yuan:2009dw}:
\begin{eqnarray}
H_1^{\perp (1) }(z) =-2zM_h  \hat H^{(3)}(z)\label{eq:hhat}
\end{eqnarray}

Thus, we can express the $x$-dependent and the $z$-dependent weighted $\sin\left(3\phi_h -\phi_S\right)$ asymmetry as
\begin{eqnarray}
A_{UT}^{Q_T^3\,\sin (3\phi_h-\phi_S)}(x) &= 2\,
\frac{\displaystyle\int dz \left[\sum\limits_a e_a^2 \, x h_{1T}^{\perp (2) a}(x)\,H_1^{\perp (1) a}(z)\right]}
     {\displaystyle\int dz \left[\sum\limits_a e_a^2\, x f_1^a(x)\,D_1^a(z)\right]} \; ,\\
A_{UT}^{Q_T^3\,\sin (3\phi_h-\phi_S)}(z) &= 2\,
\frac{\displaystyle\int dx \left[\sum\limits_a e_a^2 \, x h_{1T}^{\perp (2) a}(x)\,H_1^{\perp (1) a}(z)\right]}
     {\displaystyle\int dx \left[\sum\limits_a e_a^2\, x f_1^a(x)\,D_1^a(z)\right]} \; .
\label{eq:asymmetry}
\end{eqnarray}

\section{NUMERICAL CALCULATION}

\label{sec:numerical}
In this section, based on the above formalism, we will present the numerical estimate for the $P_{hT}^3$-weighted $\sin\left(3\phi_h -\phi_S\right)$ asymmetry in charged pion production SIDIS process at the kinematics regions of EIC and EicC.

In order to obtain the result of the denominator in Eq.~(\ref{eq:asymmetry}),
the collinear unpolarized distribution function $f_1^q(x)$ and the collinear unpolarized fragmentation function $D_1^q(z)$ are needed as the input.
For the proton distribution $f_1(x)$, we adopt the extraction from the GRV98LO PDF set~\cite{Gluck:1998xa}, while
for the fragmentation function $D_1(z)$ we apply the DSS parametrization~\cite{deFlorian:2007aj}.

As for the numerator in Eq.~(\ref{eq:asymmetry}), one needs the information of the second transverse-moment of proton pretzelosity distribution function and the first transverse-moment of the pion Collins function.
In Ref.~\cite{Lefky:2014eia}, the authors have extracted pretzelosity for the first time by using the preliminary data from COMPASS~\cite{Parsamyan:2007ju,Parsamyan:2013fia}, HERMES~\cite{Diefenthaler:2010zz,Schnell:2010zza,Pappalardo:2010zz}, and JLab~\cite{Zhang:2013dow} on $\sin\left(3\phi_h -\phi_S\right)$ asymmetry.
Assuming a Gaussian form for the $\bm p_T$-dependence for $h_{1T}^\perp(x,\bm{p}_T^2)$ , the parametrization was proposed as follows~\cite{Lefky:2014eia}
\begin{eqnarray}
h_{1T}^{\perp a}(x,\bm p_{T}^2) = \frac{M^2}{M_T^2} e^{-\bm p_T ^2/M_T^2} h^{\perp a}_{1T}(x) \frac{1}{\pi  \langle  p_T^2\rangle }\,\exp\left(-\frac{\bm{p}_{T}^2}{ \langle p_T^2 \rangle}\right)\;,
\label{eq:h1Tperp}
\end{eqnarray}
where $\langle p_T^2 \rangle =0.25$ GeV$^2$ is the mean value of $\bm p_T^2$, and
\begin{eqnarray}
h^{\perp a}_{1T}(x) &=& e  {\cal N}^a(x) (f_{1}^{a}(x) - g_{1}^{a}(x)), \label{eq:hx_par}\\
{\cal N}^a(x) &=& N^{a} x^{\alpha} (1-x)^{\beta} \frac{(\alpha + \beta)^{\alpha + \beta}}{\alpha^{\alpha} \beta^{\beta}}\, ,
\end{eqnarray}
and the values of the parameters $\alpha$, $\beta$, ${N}^a$ and $M_T^2$ obtained in the fit are ~\cite{Lefky:2014eia}
\begin{eqnarray}
 \alpha &=& 2.5\pm1.5, \quad \quad  \beta = 2~(\mathrm{fixed})  \nonumber\\
 N_{u} &=& 1 \pm 1.4,\ \quad \quad N_{d} = -1 \pm 1.3,  \quad \quad M_T^2 = 0.18 \pm  0.7 (\mathrm{GeV^2}).
 \label{eq:fit}
\end{eqnarray}
Here, $g_{1}(x)$ is the helicity distribution, for which the parametrization from Ref.~\cite{deFlorian:2009vb} is adopted.
Thus, $h_{1T}^{\perp (2) a}(x)$ derived from Eq.~(\ref{eq:moment}) has the form
\begin{eqnarray}
h_{1T}^{\perp (2) a}(x) = \frac{h^{\perp a}_{1T}(x) (M_T^2)^2 (\langle p_T^2\rangle)^2 }{2M^2 (M_T^2 +  \langle p_T^2\rangle )^3} \, .
\end{eqnarray}
Here, we point out that there are still large errors in the parametrization of the proton pretzelosity distribution function due to the limited amount of the experimental data.

In the literature there are intensive studies on the extraction~\cite{Anselmino:2007fs,Anselmino:2008jk,Anselmino:2013vqa,Anselmino:2015sxa,Anselmino:2015fty} of the Collins function using the SIDIS and the electron-positron annihilation data.
In Ref.~\cite{Kang:2015msa}, the T-odd fragmentation function $\hat H^{(3)}(z,Q)$, which the twist-3 collinear counterpart of the Collins function, is parameterized at the initial scale $Q_0= 2.4$ GeV as follows,
\begin{eqnarray}
\hat{H}_{fav}^{(3)}(z,Q_0)&= N_{u}^c z^{\alpha_{u}}(1-z)^{\beta_{u}} D_{\pi^+/u}(z,Q_0) \ , \nonumber\\
\hat{H}_{unf}^{(3)}(z,Q_0)&= N_{d}^c z^{\alpha_{d}}(1-z)^{\beta_{d}} D_{\pi^+/d}(z,Q_0) \ , \label{eq:para} \\
\hat{H}_{unf s}^{(3)}(z,Q_0) &=  N_{d}^c z^{\alpha_{d}}(1-z)^{\beta_{d}} D_{\pi^+/s, \bar s}(z,Q_0) \ ,\nonumber
\end{eqnarray}
where $\hat{H}_{fav}^{(3)}$, $\hat{H}_{unf}^{(3)}$, $\hat{H}_{unf s}^{(3)}$ represent the favored, unfavored, and strange quark unfavored fragmentation functions, respectively.
This parametrization were applied to perform a global fit, combined with the parametrization of the transversity, from the $e^+e^-\to h_1 h_2 X$ data measured by Belle and BABAR Collaborations, and the SIDIS data from HERMES, COMPASS, and JLab HALL A experiments.
As $\hat H^{(3)}$ is directly related to the first transverse-moment of the Collins function used in our calculation (see Eq.~(\ref{eq:hhat})), in this work we apply the parametrization of the Collins function $\hat H^{(3)}$  from Ref.~\cite{Kang:2015msa}.
The extracted values of the parameters in global fit are~\cite{Kang:2015msa}:
\begin{eqnarray}
N_u^c &= -0.262\pm 0.025 \quad \quad  \alpha_u =  1.69 \pm 0.01  \quad \quad \beta_u =  0.00 \pm 0.54 \nonumber\\
N_d^c &= 0.195\pm 0.007 \quad \quad  \alpha_d =  0.32 \pm 0.04 \quad \quad \beta_d =  0.00 \pm 0.79.
\end{eqnarray}

Since the parametrizations was performed at a fixed energy scale, and the kinematics of the experimental facility at EIC and EicC cover a wide range of energy, the evolution of the distribution function and the fragmentation function should be taken into account.
Thus, we customize the evolution package {\sc HOPPET}~\cite{Salam:2008qg} to include the evolution kernels for $H_1^{\perp(1)}$ and $h_{1T}^{\perp(2)}$.
The evolution kernel is adopted as the same as that of the transversity $h_1$, which has also been chosen in Ref.~\cite{Kang:2015msa,Xue:2020xba}.

\begin{figure}
  \centering

  \includegraphics[width=0.42\columnwidth]{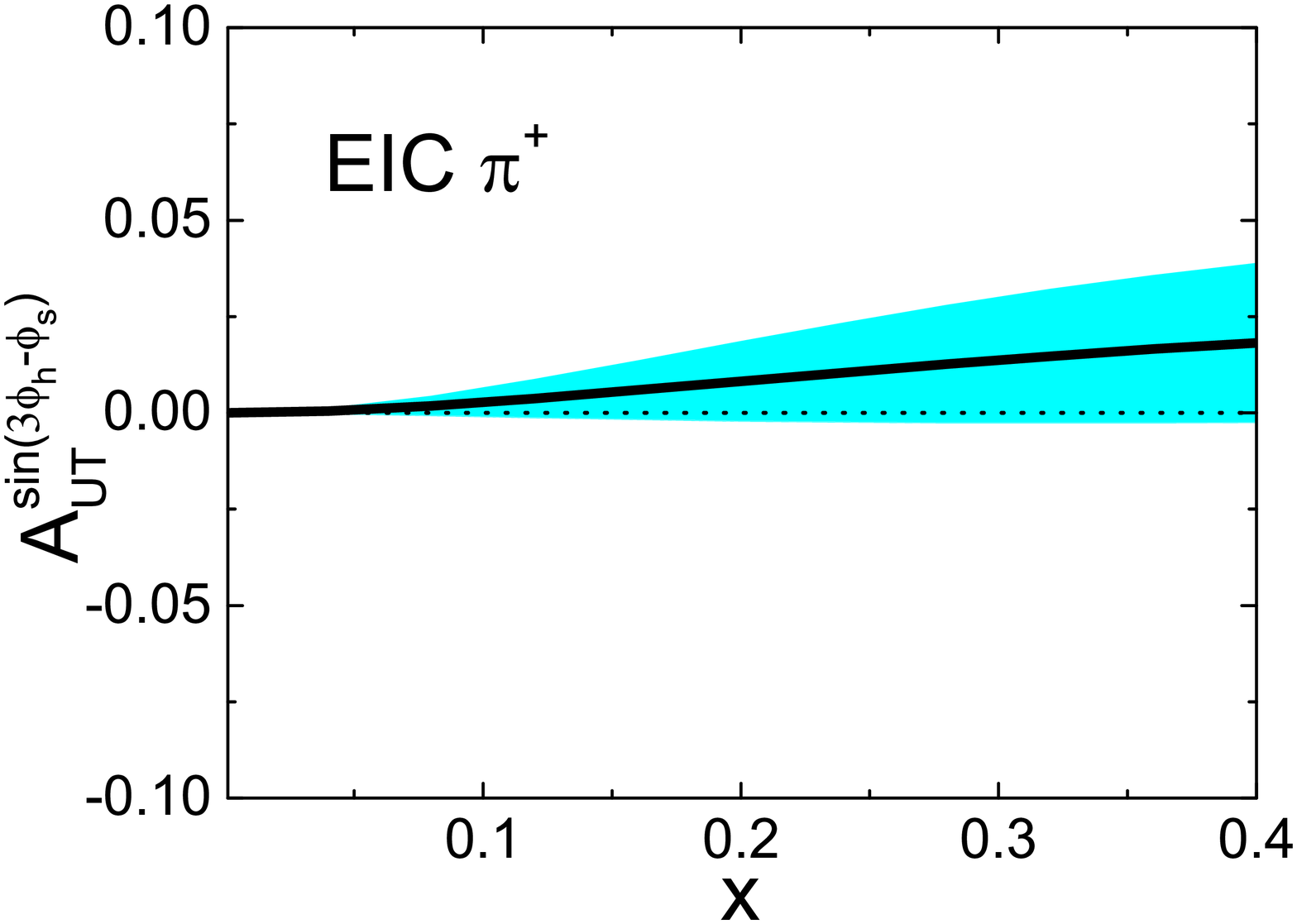}
  \includegraphics[width=0.42\columnwidth]{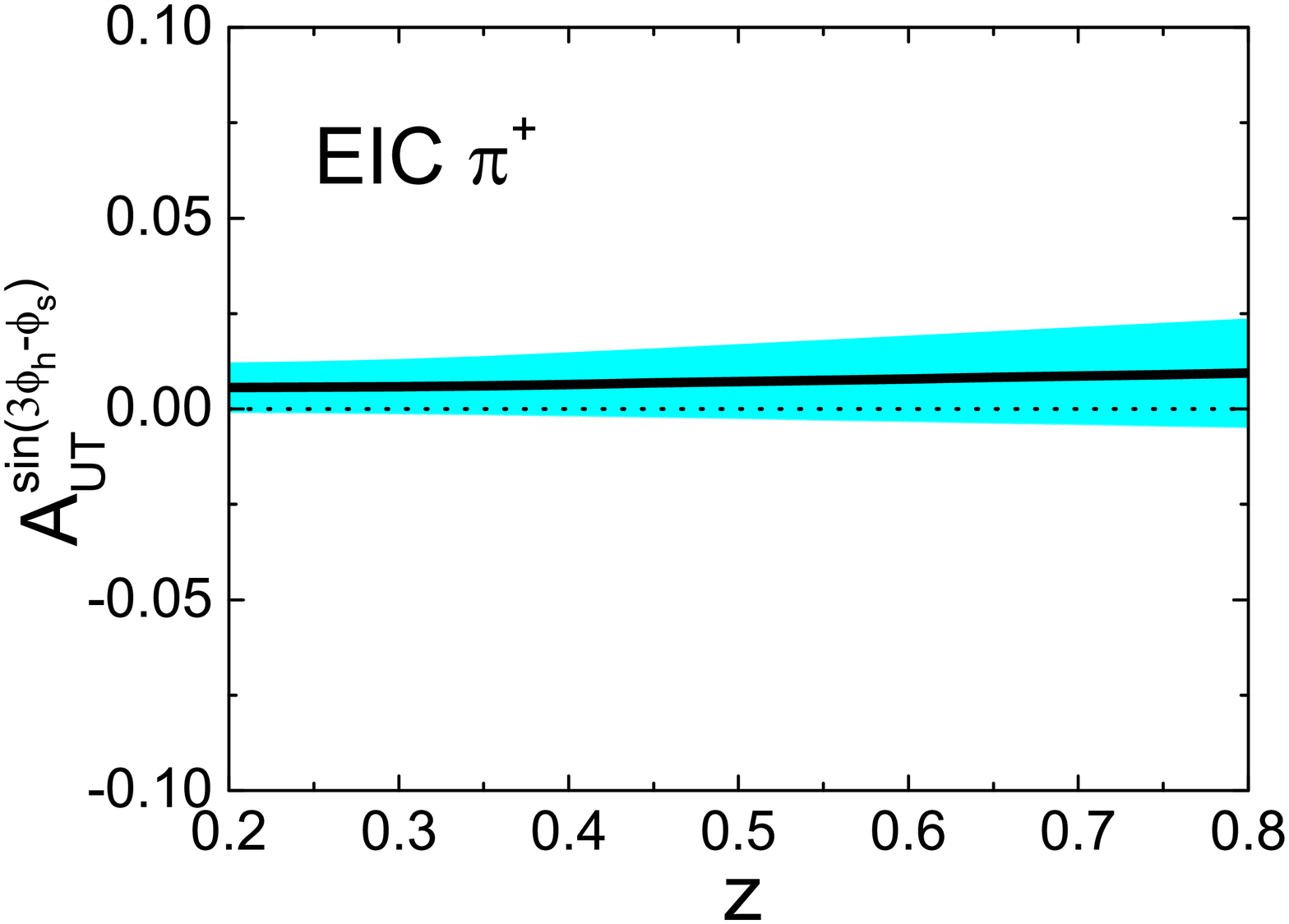}
  \includegraphics[width=0.42\columnwidth]{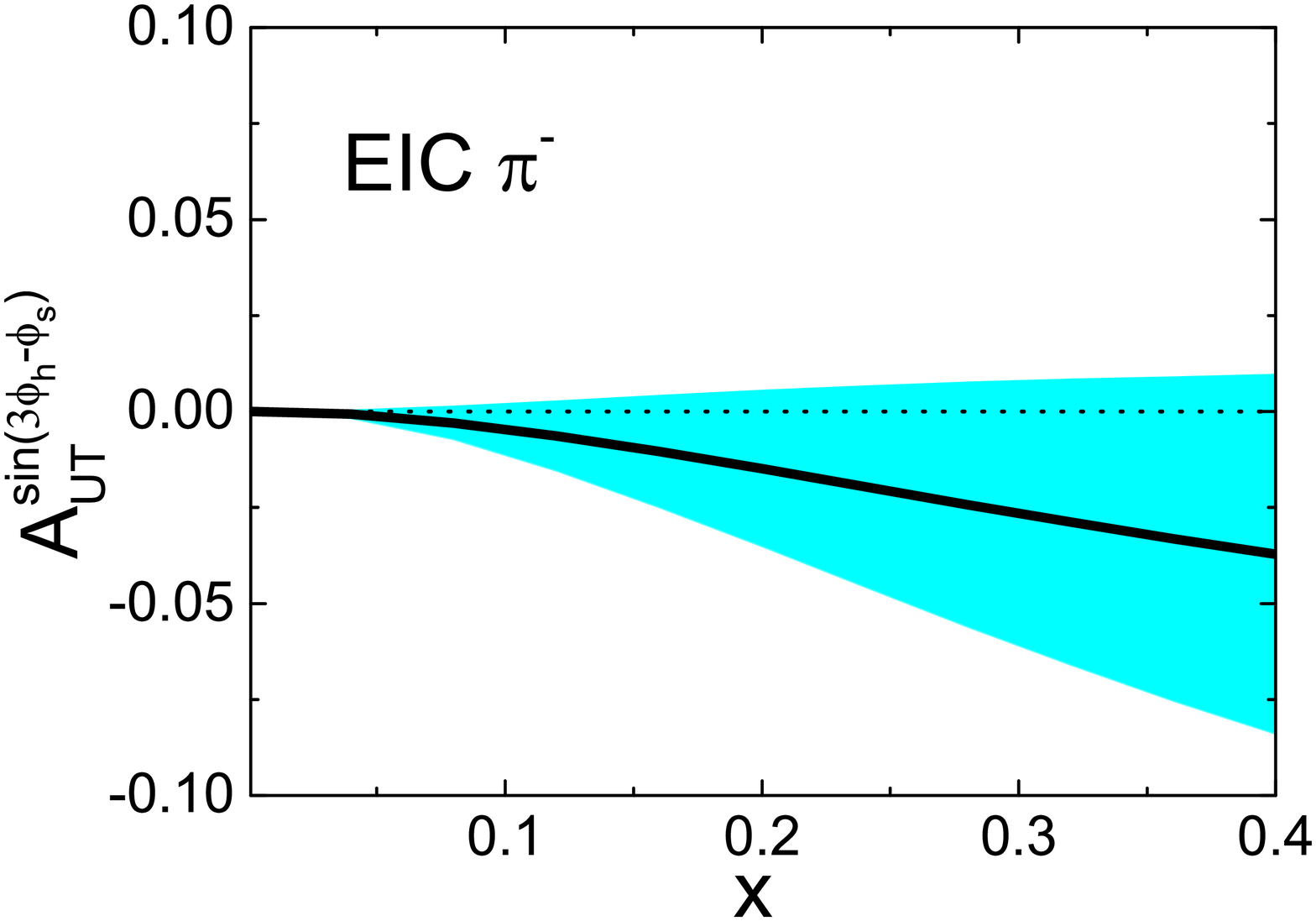}
  \includegraphics[width=0.42\columnwidth]{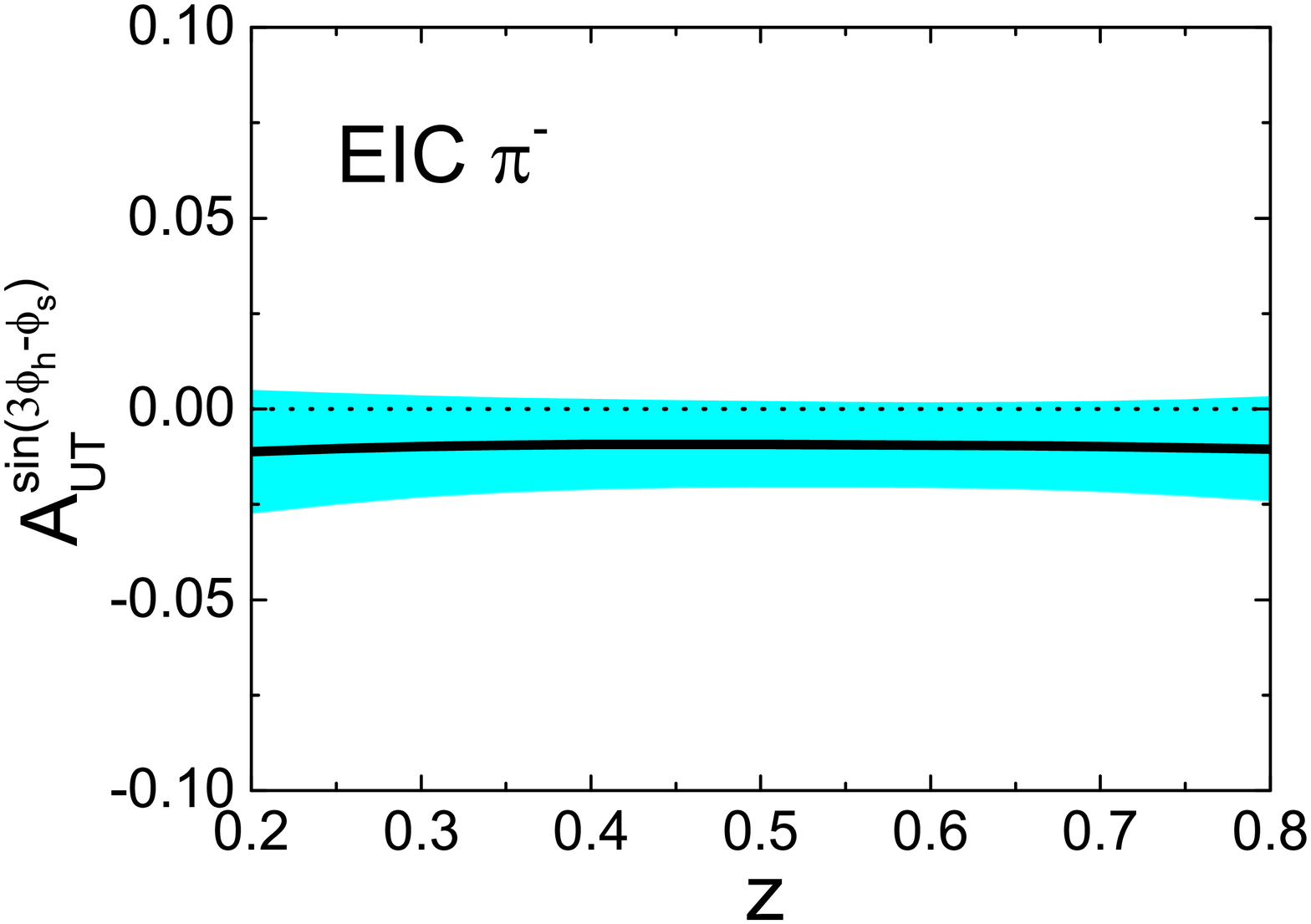}
  \includegraphics[width=0.42\columnwidth]{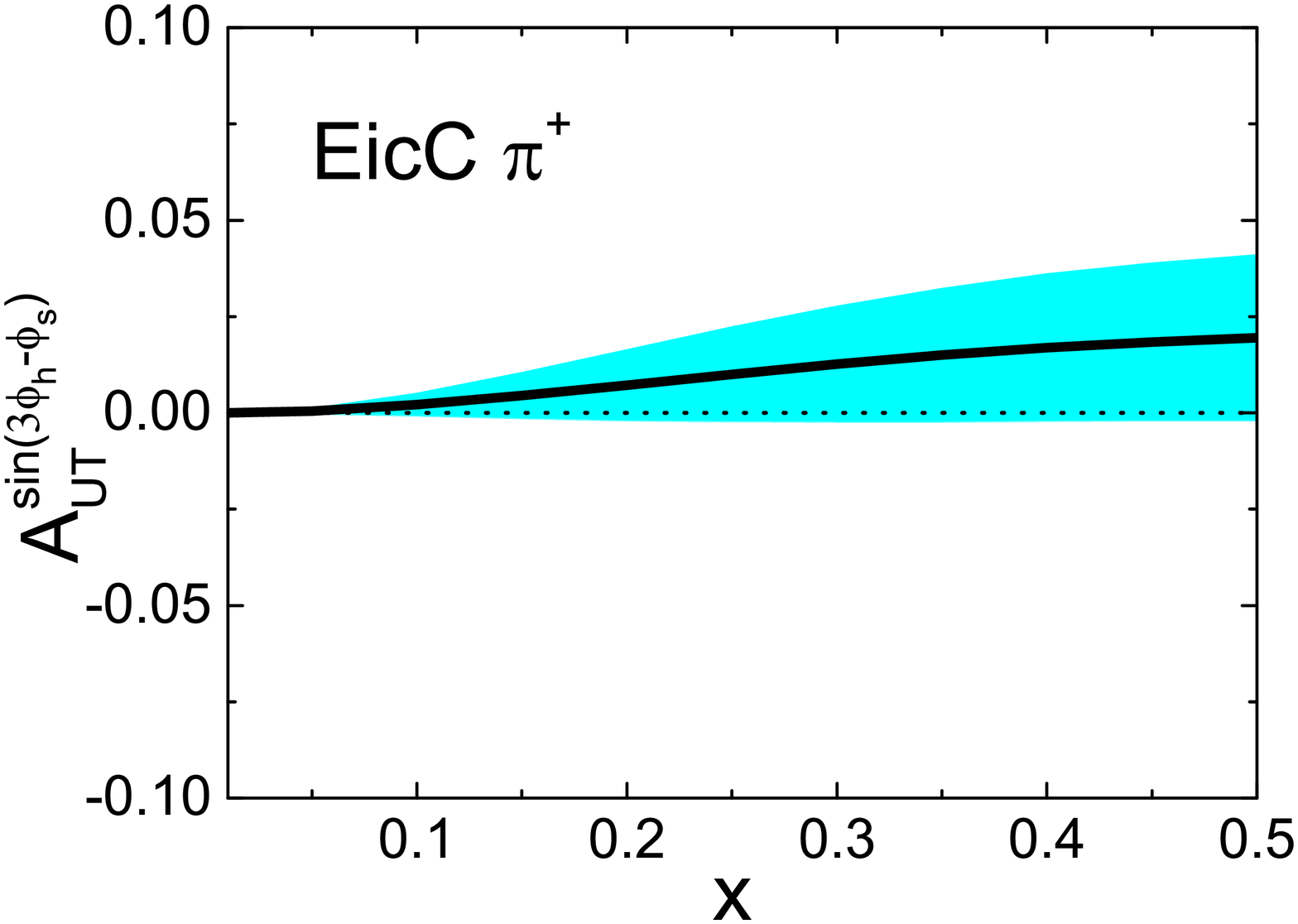}
  \includegraphics[width=0.42\columnwidth]{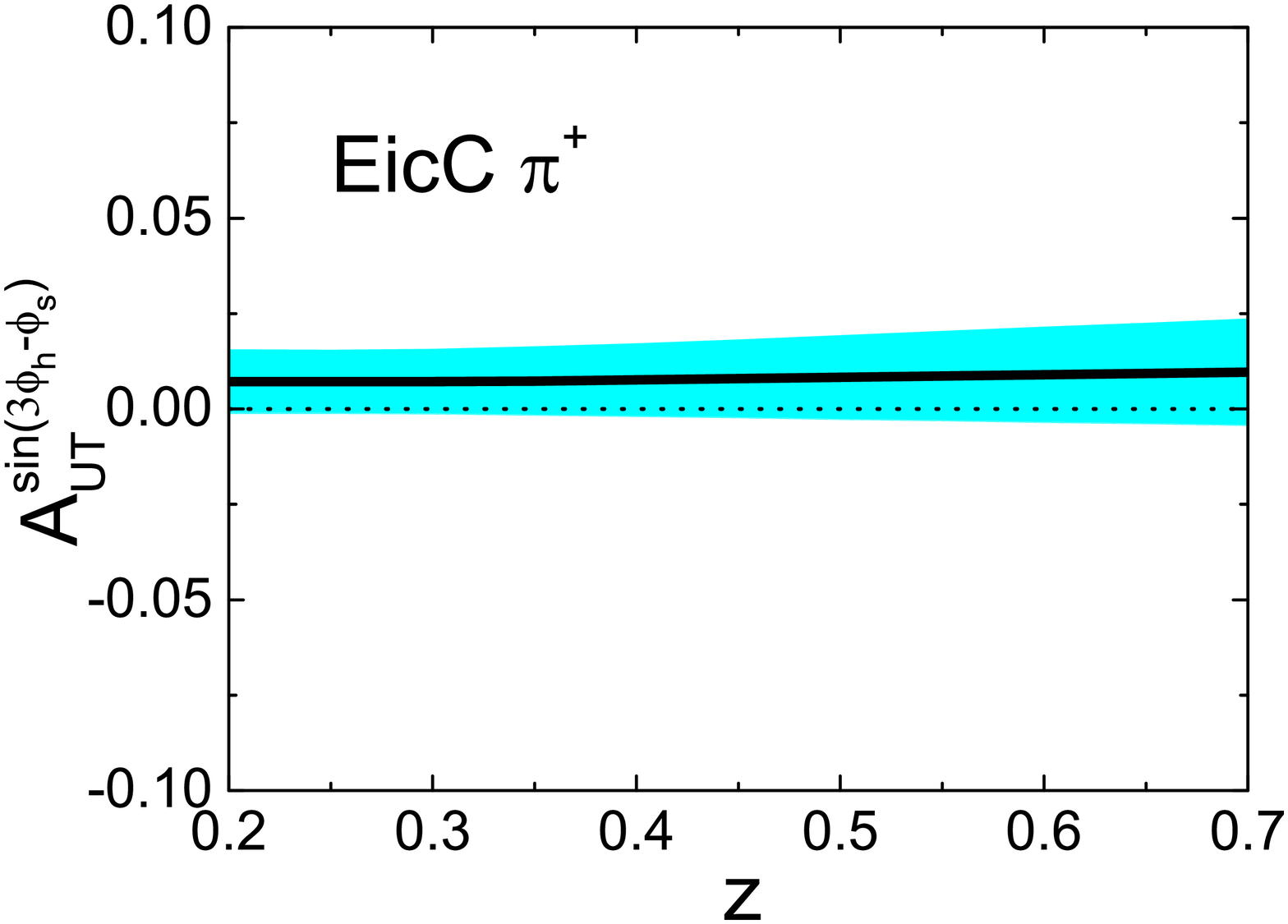}
   \includegraphics[width=0.42\columnwidth]{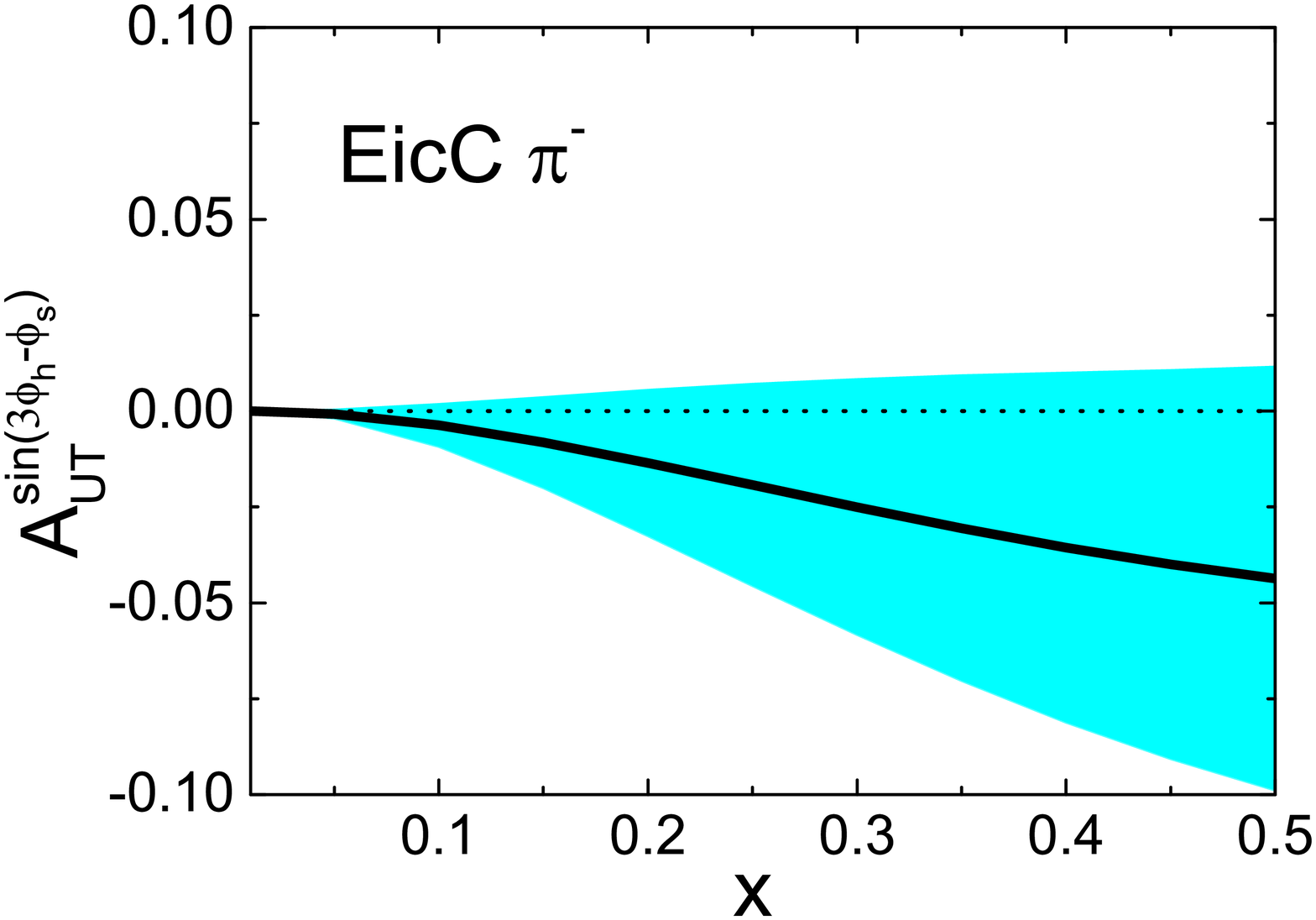}
  \includegraphics[width=0.42\columnwidth]{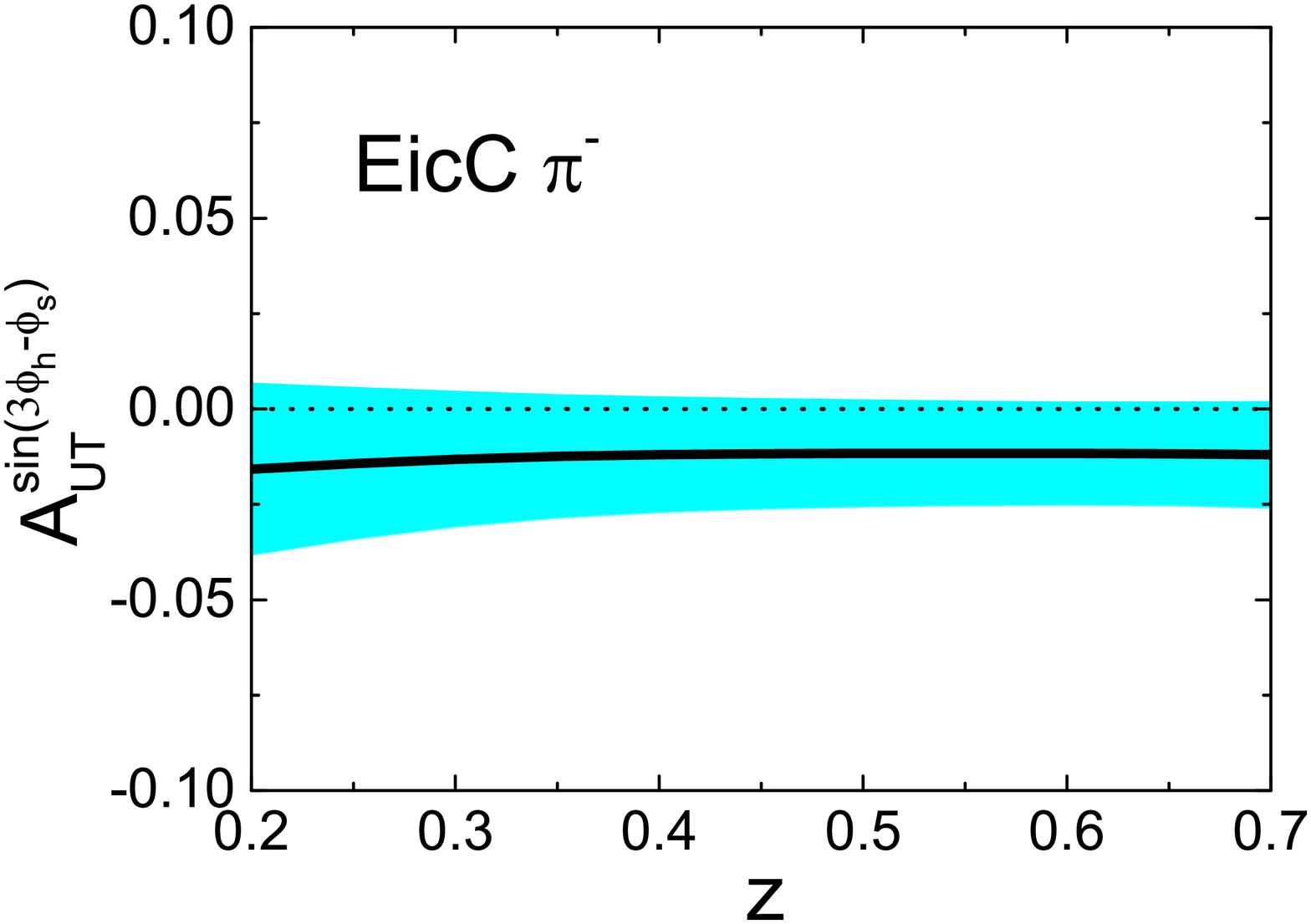}
   \caption{The $\bm{P}_{hT}^3$-weighted $\sin\left(3\phi_h -\phi_S\right)$ asymmetry at EIC and EicC as functions of x(left panels) and z(right panels).
  The solid lines correspond to the results from the central values of the parameters,
  while the shaded area show the uncertainty bands determined by the uncertainty of pretzelosity in Eq.~(\ref{eq:fit}).}
  \label{fig:asy}
\end{figure}

For the kinematical region that is available at EIC, our choices are as follows \cite{Accardi:2012qut}
\begin{eqnarray}
&0.001<x<0.4,\quad 0.1<y<0.95, \quad 0.2<z<0.8, \nonumber    \\
& 1\ \mathrm{GeV}^2<Q^{2} \;, \quad   W>5\ \mathrm{GeV},\quad \sqrt{s}=100 \ \mathrm{GeV}.
\end{eqnarray}
As for the EicC, we adopt the following kinematical cuts~\cite{Cao:2020,Anderle:2021wcy,Xue:2020xba}
\begin{eqnarray}
&0.01<x<0.5, \quad 0.3<y<0.9, \quad 0.2<z<0.7,  \nonumber    \\
 &1 \mathrm{GeV}^2<Q^2<200\ \mathrm{GeV}^2 \;, \quad   W>2\ \mathrm{GeV},\quad \sqrt{s}=16.7 \ \mathrm{GeV},
\end{eqnarray}
where $W^2=(P+q)^2\approx \frac{1-x}{x}Q^2$  is the invariant mass squared of the virtual photon-nucleon system.

Utilizing the above kinematical configurations and applying Eqs.~(\ref{eq:asymmetry}),
we numerically calculate the weighted $\sin\left(3\phi_h -\phi_S\right)$ asymmetry of pion production in SIDIS process at EIC and EicC.
The corresponding results are plotted in Fig.~\ref{fig:asy},
The upper four figures show the prediction on the $P_{hT}^3$-weighted $\sin\left(3\phi_h -\phi_S\right)$ asymmetry at EIC for $\pi^+$ and $\pi^-$ production,
while the lower four figures plot the results at EicC.
The left and right panels show the weighted $\sin\left(3\phi_h -\phi_S\right)$ asymmetry as functions of $x$ and $z$, respectively.
In each figure, the solid lines correspond to the results from the central values of the parameters in the parametrization of $h_{1T}^\perp$ and $\hat H^{(3)}$,
while the shaded area represents the uncertainty due to the statistical uncertainty in the parametrization of the pretzelosity in Eq.~(\ref{eq:fit}).
Our numerical estimates show that the $P_{hT}^3$-weighted $\sin\left(3\phi_h -\phi_S\right)$ asymmetry is around $2$ percent, and the results at the EIC are consistent with those at the EicC.
Generally, the size of the $x$-dependent asymmetry is larger than that of the $z$-dependent asymmetry.
The sign of the asymmetry is positive in the production of $\pi^+$, while it is negative for the case of $\pi^-$.
We find that the magnitude of the $x$-dependent weighted asymmetry increases with increasing $x$,
while the $z$-dependence is almost flat.
For the $x$-dependent asymmetry, the magnitudes in the production of $\pi^-$ are about twice as large as those in $\pi^+$ both at EicC and EIC.
This can be explained by the fact that the size of pretzelosity of the down quark is larger than that of the up quark in the extraction in Ref.~\cite{Lefky:2014eia}, and in the $\pi^-$ production process the down quark dominates.
Due to the large uncertainties in the parametrization of pretzelosity, the uncertainty bands of the weighted asymmetry are wide, particularly in the larger $x$ region.
The future measurements on the $P_{hT}^3$-weighted $\sin\left(3\phi_h -\phi_S\right)$ asymmetry in SIDIS at EIC and EicC with high precision could provide more stringent constraint on
$h_{1T}^{\perp (2)}$ and reduce the uncertainty in a wide $x$ range.

\section{CONCLUSION}
\label{sec:conclusion}

In this work, we have studied the $P_{hT}^3$-weighted asymmetry with a $\sin\left(3\phi_h -\phi_S\right)$  modulation in charged pion production in SIDIS process.
The asymmetry is contributed by the convolution of the second transverse-moment of pretzelosity and the first transverse-moment of the Collins function, for which we have utilized recent parametrizations from literature.
We have taken into account the DGLAP evolution effects for both the collinear distribution functions and the fragmentation functions.
To do this we have customized the package {\sc{HOPPET}} to include the evolution kernel for the corresponding distributions and fragmentation functions.
The $P_{hT}^3$-weighted $\sin\left(3\phi_h -\phi_S\right)$ asymmetries in charged pion production in SIDIS process at the kinematics configurations of EIC and EicC have been estimated.
The numerical result shows that the weighted asymmetry is around several percents, in both EIC and EicC.
The sign of the asymmetry is generally positive in the production of $\pi^+$, while it is negative for the case of $\pi^-$.
Furthermore, the uncertainty of the asymmetry due to the uncertainty of the parametrization for pretzelosity have also presented and is rather large in the larger $x$ region.
Our study shows that, using $H_1^{\perp (1)}(z)$ as a probe, the $P_{hT}^3$-weighted $\sin\left(3\phi_h -\phi_S\right)$ asymmetry at EICs can be a useful tool to access the second transverse moment of pretzelosity as well as to provide further constraints on pretzelosity.

\section{ACKNOWLEDGMENTS}
This work is partially supported by the NSFC (China) grants 11905187,11847217,11575043 and 11120101004. X. Wang is supported by the China Postdoctoral Science Foundation under Grant No.~2018M640680 and the Academic Improvement Project of Zhengzhou University.

\end{document}